\documentclass[aps,prl,reprint,groupedaddress]{revtex4-1}

\usepackage{graphicx}
\usepackage{bm}
\usepackage{hyperref}
\usepackage{listings}
\usepackage{color}
\usepackage{caption}
\usepackage{subcaption}
\usepackage{mathtools}
\usepackage{cleveref}
\usepackage{pgfplots}
\usepackage{tikz}
\usetikzlibrary{plotmarks,backgrounds}
\usepackage{amssymb}
\pgfplotsset{compat=newest}
\pgfplotsset{plot coordinates/math parser=false}

\usepgfplotslibrary{external} 
\newlength\fwidth
	\newlength\fheight
	
\newcommand{\vb}{\textbf}
\newcommand{\tens}[1]{\underline{\underline{#1}}}
	
\begin{document}
\title{Magnetic reconnection in plasma under inertial confinement fusion conditions driven by heat flux effects in Ohm's law}
\author{A.S. Joglekar} 
\affiliation{Department of Nuclear Engineering and Radiological Sciences,  Center for Ultrafast Optical Science, Ann Arbor, Michigan 48109, USA}
\author{A.G.R. Thomas}
\email{agrt@umich.edu}
\affiliation{Department of Nuclear Engineering and Radiological Sciences,  Center for Ultrafast Optical Science, Ann Arbor, Michigan 48109, USA}
\author{W. Fox}
\affiliation{Princeton Plasma Physics Laboratory, P.O. Box 451, Princeton, New Jersey 08543, USA}
\author{A. Bhattacharjee}
\affiliation{Princeton Plasma Physics Laboratory, P.O. Box 451, Princeton, New Jersey 08543, USA}
\date{\today}

\begin{abstract}
In the interaction of high-power laser beams with solid density plasma there are a number of   mechanisms that generate  strong magnetic fields. Such fields  subsequently inhibit or redirect electron flows, but can themselves be advected by heat fluxes, resulting in complex interplay between thermal transport and  magnetic fields. We show that for  heating by multiple laser spots reconnection of magnetic field lines can occur, mediated by  these heat fluxes, using a fully implicit 2D Vlasov-Fokker-Planck code. Under such conditions, the reconnection rate is dictated by heat flows rather than  Alfv\`enic flows. We find that this mechanism is only relevant in a high $\beta$ plasma. However, the Hall parameter $\omega_c\tau_{ei}$ can be  large so that thermal transport is strongly modified by these magnetic fields, which can impact  longer time scale temperature homogeneity and ion dynamics in the system.
\end{abstract}
\maketitle
Understanding the $\mathcal{O}\left(10^2\right)$ T magnetic fields that can develop in high-power-laser interactions with solid-density plasma  \cite{PhysRevLett.26.1012,PhysRevLett.80.5137,PhysRevLett.88.045004,Tatarakis:2002fk,PhysRevE.70.026401} is important because such fields significantly modify both the magnitude and direction of electron heat fluxes \cite{1965RvPP....1..205B}.  The dynamics of such fields evidently has consequences for inertial fusion energy applications \cite{ISI:000253530600037,ISI:000079969500054,Li_Sci_2010}, as the coupling of the laser beams with the walls or pellet and the development of temperature inhomogeneities  are  critical to the uniformity of the implosion. There is a significant interplay between heat fluxes and magnetic fields:  in semi-collisional plasmas heat flux can be the dominant mechanism for transporting magnetic fields in addition to currents or bulk ion flow \cite{ISI:A1986F164400007}. This effect, arising due to an electric field analogous to the Nernst-Ettingshausen effect in metals \cite{ANDP:ANDP18862651010}, has been shown to be significant  in laser heated plasma \cite{ISI:A1984SZ82700013,ISI:000253336900040,ISI:000281164200010}. The Nernst effect in plasma \cite{ISI:A1986F164400007} arises as a consequence of the velocity dependent collision frequency of electrons in plasma. Since the faster, ``hot'', population of electrons are essentially collisionless, the magnetic field is ``frozen'' to them, whereas the collisional, ``cold'', portion of the distribution function is able to diffuse across field lines. Hence, magnetic fields can be advected with close to zero net current by ``hot'' electrons.

In heating plasma with a finite laser spot, an azimuthal magnetic field about the heated region arises through the Biermann battery effect \cite{ISI:A1950UW42200001}. For multiple spots in close proximity, as in inertial fusion, these magnetic fields will be in a configuration with oppositely directed field lines. Under such conditions, magnetic reconnection of field lines may be expected to arise. 
Magnetic reconnection has been intensely studied in space plasmas, but more recently, laser inertial fusion relevant scenarios  have been investigated \cite{ISI:000243414600032,Li_PRL_2007,Willingale_POP_2010}.

 In Sweet-Parker theory, a resistive region  between plasma inflows with resistivity $\eta$  allows magnetic field lines to diffuse and change topology, leading to jets outflowing at Alfv\`enic speeds, $v_A$ \cite{ISI:A19679100300028,ISI:A1972M695900005}. However, observed reconnection rates are rarely well described by this model.
In case of a small diffusion region $L$ for which the Sweet-Parker width, $\delta_{SP} = L/\sqrt{S}$ where $S=v_AL\mu_0/\eta$ is the Lunquist number, 
is smaller than the ion inertial length, $c/\omega_{pi}$, Hall physics is relevant and reconnection is no longer resistivity dominated.  
Rather, augmented by the inclusion of Hall physics,  reconnection rates have been shown to be significantly faster suggesting that the dynamics at such small scale lengths contribute strongly to reconnection. Recently, researchers have been interested in the intermediate regime between collisionless and collisional reconnection \cite{Cassak_PRL_2005,ISI:000268615200019,ISI:000291018000004,Fox_POP_2012}, where the Nernst effect is important. 
Given the analogous form of the Nernst term and Hall current term in Ohm's law, it is natural to assume that the Nernst effect may enable reconnection in a similar manner to Hall reconnection, but with the electron currents replaced by heat fluxes.

In this Letter, we show that under conditions similar to those found in hohlraums, where heat flux effects in Ohm's law are important \cite{ISI:A1986F164400007}, reconnection of field lines can occur. The heat fluxes that are generated by the laser hot-spots drive reconnection through advection at the ``Nernst'' velocity $v_T$. The Nernst effect allows magnetic field advection without an associated electron current, which is different than the standard Hall effect within the reconnection layer; this breaks the Alfvenic constraint (at least within the parameters considered) and allows characteristic reconnection rates of $E_z/(B_0v_T)$ rather than $E_z/(B_0v_A)$. We show that this can occur for conditions described by a dimensionless number describing the ratio of Nernst to  electron flow velocities. We find that this mechanism is only relevant in a high $\beta$ plasma, i.e. where the ratio of thermal pressure to magnetic pressure is large. However, the Hall parameter $\omega_c\tau_{ei}$ can simultaneously be  large so that thermal transport is strongly modified by  magnetic fields, which can impact  longer time scale temperature homogeneity and ion dynamics.

The Vlasov-Fokker-Planck (VFP) equation,
is solved together with Ampere's and Faraday's Laws to describe the plasma.
The code we use, \textsc{Impacta} \cite{ISI:000189117600001,ISI:000263744500001}, uses a Cartesian tensor expansion \cite{ISI:A1960WC10700004}, with the distribution function expanded as $f(t,\vb{r},\vb{v})=f_0 + \vb{f}_1\cdot\hat{\vb{v}} + \tens{f}_2:\hat{\vb{v}} \hat{\vb{v}} +\dots$ This expansion can be truncated in a collisional plasma, as collisions tend to smooth out angular variations in the  distribution function, resulting in a close to isotropic distribution in the centre of mass frame, represented by $f_0$. Higher orders are successively smaller perturbations, $\tens{f_2} \ll \vb{f}_1 \ll f_0$ etc.
Using the Lorentz gas approximations, electron-ion collisions appear in the equation describing the evolution of $\vb{f}_1$ in the ion center-of-momentum frame  as an effective collision frequency $\propto 1/v^3$:
 \begin{eqnarray}
  \frac{\partial{{\bf f}_1}}{\partial{t}}
              & + & {v}{\nabla}{f}_0 
	        - \frac{e\textbf{E}}{m_e}\frac{\partial{f}_0}{\partial{v}} 
		-  \frac{e\textbf{B}}{m_e}\times{\bf {f}}_1
                + \frac{2}{5}{v}\:{\nabla}\cdot\tens{{f}}_2 \nonumber\\  
			&-& \frac{2}{5{v}^3}\:\frac{\partial}{\partial{v}}
			\left({v}^3\frac{e\textbf{E}}{m_e}\cdot\tens{{f}}_2\right)
		= -\frac{Yn_iZ^2}{v^3}{\bf {f}}_1
		 \label{f1}
\end{eqnarray}
 where $Y=4\pi(e^2/4\pi\epsilon_0m_e)^2\ln\Lambda_{ei}$. In \textsc{Impacta}, terms up to and including $\tens{f}_2$ are retained. 
 
 In reconnection studies, Ohm's law is of crucial significance. We can formulate a generalized form of Ohm's law for this velocity dependent collision operator by multiplying by $v^3$ and taking the current moment $(4\pi/3)\int_0^\infty \dots v^3 dv$. 
\begin{multline}
\textbf{E}
= \overline{\eta} \textbf{j}  +\frac{\vb j \times \vb B}{e n_e} -  \vb v_T \times \vb B\\
-\frac{\nabla\left( n_e m_e \langle v^5\rangle\right)}{6e n_e \langle v^3\rangle}- \frac{\nabla\cdot\left(n_e m_e \langle \vb v \vb v v^3\rangle\right)}{2 e n_e \langle v^3\rangle }\;,
\label{ohmv6}
\end{multline}
where the effective resistivity is $$ \overline{\eta}=\frac{2\pi Z e^2 \ln \Lambda_{ei}}{(4\pi\epsilon_0)^2m_e \langle v^3\rangle }\;,$$ the  magnetic convection velocity  by heat flow  \cite{ISI:A1986F164400007} is $$
\vb{v}_T = \frac{\langle\vb v v^3\rangle }{2 \langle v^3\rangle }+\frac{\vb{j}}{e n_e}\;,
$$
 the inertial term ($\partial/\partial t$) is neglected, valid for a sufficiently collisional system, a term contracting with $\vb{E}$, $ {\langle\vb v \vb v v\rangle}/(2{\langle v^3\rangle})$, is assumed to be small,  and velocity moments are defined by
\begin{eqnarray}
\langle v^n\rangle  &= \frac{4 \pi}{n_e} \int_0^\infty f_0 ~ v^{n+2} dv\;, \nonumber\\
\langle \vb v v^n\rangle  &= \frac{4 \pi}{3 n_e} \int_0^\infty \vb{f}_1 ~ v^{n+3} dv\;, \nonumber\\
\langle \vb v \vb v v^n\rangle  &= \frac{8 \pi}{15 n_e} \int_0^\infty \tens{f_2} ~ v^{n+4} dv\;. \nonumber
\end{eqnarray}
The last two terms in Eqn.  \ref{ohmv6} combined play the role of the pressure tensor term normally used in Ohm's law. To express Eqn.  \ref{ohmv6} in a more familiar form, if we assume that the distribution function is a Maxwellian speed distribution multiplied by a function of angle only, 
 it can be shown that Eqn.  \ref{ohmv6} reduces to 
\begin{multline}
\textbf{E}= \overline{\eta} \textbf{j}  +\frac{\vb j \times \vb B}{e n_e}-\frac{\nabla\cdot\tens{P}_e}{e n_e}-  \vb v_T \times \vb B-\frac{3}{2}\frac{\nabla T_e}{e}\;,
\label{ohmv7}
\end{multline}
where $\tens{P}_e$ is the full electron pressure tensor and we have neglected a $ \sim{\langle\vb v \vb v \rangle}/{\langle v^2\rangle}$ correction to the $\nabla T_e$ term. We do not use Eqn.  \ref{ohmv7} in this study, but instead compare the results from the Vlasov-Fokker-Planck code with the more general Eqn.  \ref{ohmv6}. 

To gain some insight into the physical meaning of $\vb{v}_T$, Haines showed, using a model $1/v^2$ collision operator  \cite{ISI:A1986F164400007}, 
 that the Nernst velocity could be related directly to the heat flux by ${\bf v}_T\simeq 2{\bf q}_e/(5p_e)$ and how it relates to terms in Braginskii's equations \cite{1965RvPP....1..205B}. 
 
To parameterize under what conditions the situation we describe may occur, we can compare the relative magnitudes of the Hall term, $\vb{j} \times \vb{B} / e n_e$, and the heat flow term, $\vb{v}_T \times \vb{B}$ in Ohm's law to generate a new dimensionless number: 
\begin{eqnarray}
H_N=\frac{en_e|\vb{v}_T|}{|\vb{j}|} = \frac{1}{5} \frac{\kappa_\perp^c}{\omega_c \tau_{ei}} \left(\frac{1}{\tilde\delta_c}\right)^2 \equiv\frac{1}{5} \kappa_\perp^c\beta\omega_c \tau_{ei}\label{dim}
\end{eqnarray}

where $\omega_c \tau_{ei}$ is the Hall parameter, $\kappa_\perp^c$ is the normalized perpendicular thermal conductivity coefficient \cite{ISI:A1986A817800020} and we have used the  heat flux component $\vb{q}_{e\perp} \sim\kappa_\perp\nabla T_e$ to estimate $\vb{v}_T$ and assumed the gradient scale lengths for the temperature and magnetic field are similar. The normalized skin depth, $\tilde\delta_c = {c}/{(v_{\text{th}}\omega_{pe} \tau_{ei})}$ serves as an  independent parameter in Eqn.  \ref{dim}. A small skin depth relative to the mean-free-path means that electron currents are inhibited, but the semi-collisional behavior still allows for electron energy transport. $\beta$ is the ratio of thermal pressure to magnetic pressure. In the limit of large $\omega_c \tau_{ei}$, the $\kappa_\perp^c$ approaches the asymptotic limit $\kappa_\perp^c=\gamma_1^\prime/(\omega_c \tau_{ei})^2$, where $\gamma_1^\prime$ is a coefficient between 3.25 and 4.66 depending on $Z$ \cite{ISI:A1986A817800020}. Hence, for large $\omega_c \tau_{ei}$, $H_N = (\gamma_1^\prime/5)\beta/\omega_c \tau_{ei}$ and can therefore only be significant for a high $\beta$ plasma.

An important parameter in magnetic reconnection is the Lundquist number $S$. We can also introduce an analogously formulated Nernst-Lundquist number, $S_N = v_TL\mu_0/\eta$, which is defined according to the usual definition, but  replacing the Alfv\`en velocity with the more relevant  Nernst velocity. The relationship between these two dimensionless parameters is 
\begin{eqnarray}
S_N = H_N\omega_c\tau_{ei} = \frac{\kappa_\perp^c}{5}\left(\frac{1}{\tilde\delta_c}\right)^2 \label{dim2}
\end{eqnarray}

From these dimensionless numbers, we can see that for an interesting heat-flux reconnection problem (i.e. for $\omega_c\tau_{ei} \geq 1$) dominated by Nernst effects ($H_N\gg1$),  the Nernst-Lundquist number must also be large, $S_N\gg1$. This means that resistive effects will be small, and therefore anisotropic pressure-like ($f_2$) effects must be included in Ohm's law to support the electric field at the $X$-point.  In ref. \cite{ISI:000268615200019}, Daughton et al included heat flux effects in their reconnection study, but for their system $H_N\lesssim 1$ so the thermal contribution was small. Here we examine a situation where $H_N\gg1$ where heat flux effects dominate.

As in ref. \cite{ISI:000189117600001,ISI:000263744500001}, we use a normalization scheme with time  normalized to $\tau_n=4\tau_{ei}/3\sqrt{\pi}$ and velocity normalized to $v_{\text{th}0} = \sqrt{2 k_B T_{e0} / m_e}$.  As a result, lengths are normalized to the electron mean-free-path $\lambda_{\text{mfp}} = v_{\text{th}}\tau_n$.  The computation is performed in a domain defined over the range $-100\lambda_{\text{mfp}} < y < 100\lambda_{\text{mfp}}$ and $-1500 \lambda_{\text{mfp}} < x < 1500 \lambda_{\text{mfp}}$. The cells near the boundary in $\hat{x}$ is exponentially increasing in step size such that they can be considered ``far away''. The domain of interest  in $\hat{x}$, where the cell size is constant, is $-400 \lambda_{\text{mfp}} < x < 400 \lambda_{\text{mfp}}$. The numerical  resolution in the runs shown in the paper is $\Delta x = 13.3333\lambda_{mfp},  \Delta y = 3.125\lambda_{mfp}, \Delta v = 0.0625v_{th}$.

The connection between the normalized quantities and  real parameters is made through the ratios $v_{\text{th}}/c$ and $\omega_{pe}\tau_{n}$. Here, $v_{\text{th}}/c = 0.08$ and $\omega_{pe} \tau_{n} = 125$ are chosen in order to put the system into inertial confinement relevant conditions, corresponding to a temperature $T_{e0} =$ 1.6 keV and electron number density $n_e = 2.5\times10^{22}~\text{cm}^{-3}$. A magnetic field of $B_0 = 1$ corresponds to a field strength of 400 T (4 MG). 

The magnetic field is generated through the $\nabla n_e \times \nabla T_e$ mechanism. We introduce an out of plane plasma density gradient of the form
${\partial n(x,y)}/{\partial z} = \frac{n_0}{L_n} e^{-(x/r_{0})^2}\left(e^{-((y+y_{\text{max}})/r_0)^2}+e^{-((y-y_{\text{max}})/r_0)^2}\right)$,  where $L_n = 50$ and $r_0 = 50$, by adding a $z$ component of electric field.  This gradient is switched off at $t=800\tau_{n}$ to prevent excessive magnetic field generation.
The temperature profile is accomplished by heating the plasma near the $y$-boundaries of the system using an inverse bremsstrahlung heating operator \cite{ISI:A1980JG47900006} with a profile 
$H(x,y) = H_0 e^{-(x/r_{0})^2}\left(e^{-((y+y_{\text{max}})/r_0)^2}+e^{-((y-y_{\text{max}})/r_0)^2}\right)$ where $H_0 = 0.5$, corresponding to a laser of intensity $2.5\times10^{14}$ Wcm$^{-2}$. 
The heated regions result in strong heat fluxes in the $\hat{y}$ direction that advect the magnetic field lines inwardly towards the reconnection region. Ions are stationary in the simulation to isolate these effects, which may be justified physically in the case of the walls of a hohlraum as they are heavy ions (for example gold) \cite{ISI:000188282100001}. Simulations run with ion motion show similar behavior. 
 A thorough study of the Nernst and bulk flow advection of magnetic fields is in Ref. \cite{ISI:000281164200010}.

\begin{figure}[h]
\includegraphics[width=\columnwidth]{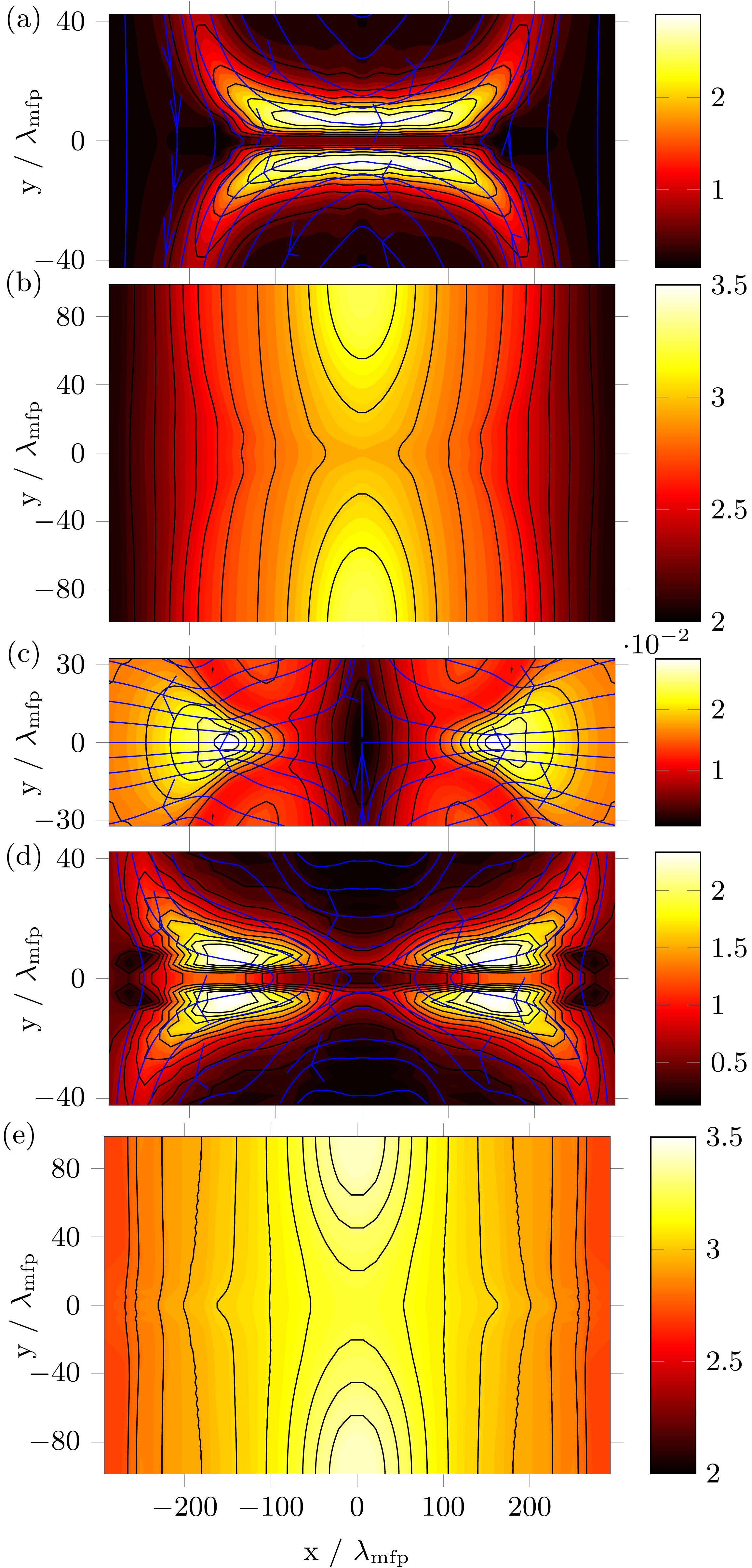}
%
\caption{At a time ${t}=19000~\tau_{n}$ into the simulation (a) 
$\vb{B}/m \nu_{ei}/e$,
(b) $T_{e}/T_{e0}$, 
(c)$ \vb{v}_{T}/ v_{\text{th}0}$ in the $x-y$ plane. 
\\ (d) $\vb{B}/m \nu_{ei}/e$ and (e) $T_{e}/T_{e0}$  at ${t}=27000~\tau_{n}$
\\ \small{Note: The axes are not square}}
\label{plots}
\end{figure}

\Cref{plots} shows output from the simulation at a time $19000~\tau_{n}$. (a) shows the magnetization of the plasma, $\vb{B}$ and (b) illustrates the temperature profile of the system. The Nernst velocity (c) is approximately $10^2$ larger in magnitude than the maximum current (not shown). The flow direction of the Nernst velocity (calculated directly from the distribution function) indicates that thermal energy is being brought inwards in the $y$-direction towards the reconnection region and is subsequently redirected outwards in the $x$-direction, carrying the magnetic field with it. Distinct ``jets'' of heat flux are formed out of the reconnection region.

\Cref{plots}(d) shows the magnetic field profile after the majority of the flux has reconnected, at a time $27000~\tau_{n}$ into the simulation, which corresponds to approximately 0.6 ns. The reconnected field lines are then advected by the Nernst jets towards the $\hat{x}$ boundaries. \Cref{plots}(e) shows the temperature profile at the same time as the magnetic field in (d). The outward heat flow in $\hat{x}$ from the reconnection process causes the change in the temperature profile from (b) to (e).

\begin{figure}[h]
\begin{center}
\includegraphics[width=\columnwidth]{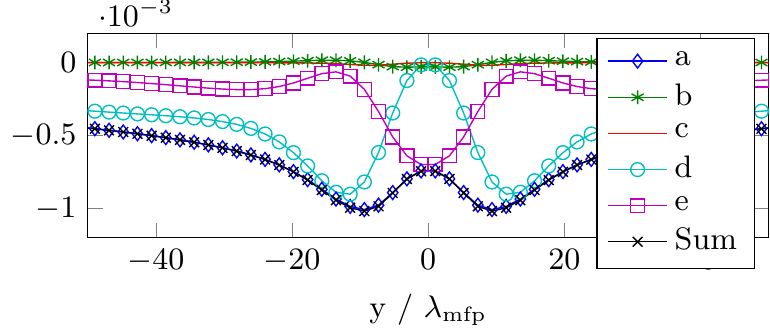}
\caption{Illustration of the contribution of the different components of Ohm's Law in \Cref{ohmv6} taken from the simulation at a time ${t}=11000~\tau_{n}$. \\(a) ${E}_{z}$ calculated from the code \\, (b) $\bar{\eta} {j}_{z}$, (c) $[\vb{j} \times \vb{B}]_{z}$, (d) $[\vb{v}_{T} \times \vb{B}]_{z}$,(e): $\left[\frac{\nabla \cdot \langle \vb{v}\vb{v} v^3\rangle }{2 \langle v^3\rangle }\right]_{z}$,\\ (f) Sum of all contributions (b-e)}
	\label{recline}
\end{center}
\end{figure}
The quantity $E_z$ is the rate at which magnetic flux crosses the neutral point. In the case of oppositely directed magnetic fields, $B_x$, the reconnecting magnetic field, $B_y$, is generated through Faraday's Law by the out of plane electric field, $\partial E_z/\partial x$ in a 2-D Cartesian geometry.   We can analyze the various contributions from the generalized Ohm's Law,  \cref{ohmv6} by directly calculating the velocity moments. \Cref{recline} shows the out-of-plane electric field, $E_z$, and four of the terms that contribute to it. Anisotropic pressure tensor-like terms almost entirely support $E_z$ at the $X$-point where the flows diverge, with a small contribution from the resistive term. The $ \vb{v}_T \times \vb{B}$ term provides  an analogue of the Hall current, with the actual Hall current  $ \vb{j} \times \vb{B}$ being negligible. The sum of just these moments of the numerical distribution function agrees well with the electric field taken from the code (which in these calculations includes electron inertia). Using the terms in \cref{ohmv7} instead, similar results are obtained, with the small difference being due to the non-Maxwellian distribution that develops in the reconnection region.  

\begin{figure}[h]
\includegraphics[width=\columnwidth]{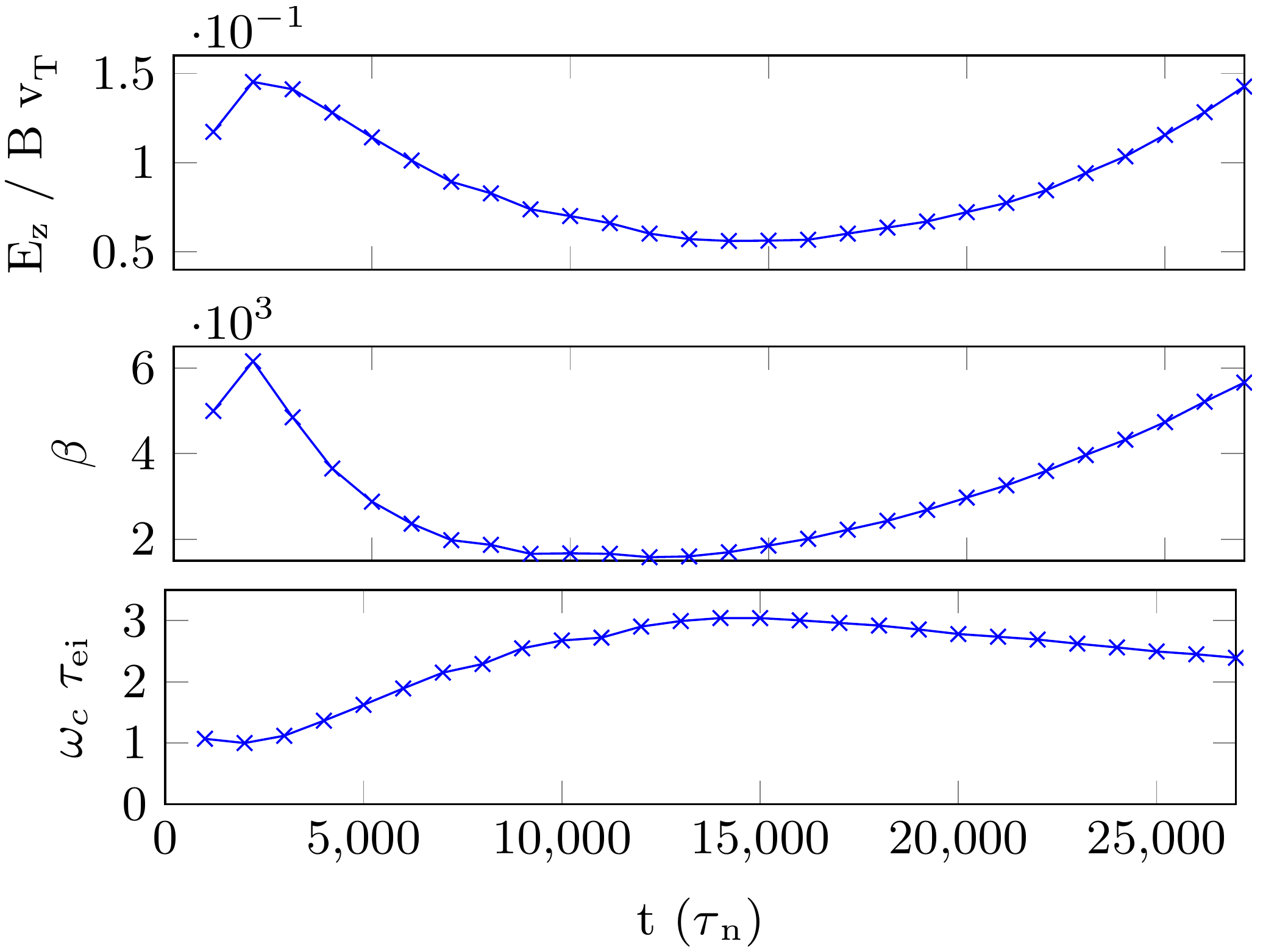}
\caption{ (a) The reconnection rate (b) $\beta$ - the ratio of thermal pressure to magnetic pressure (c) $\omega_c \tau_{ei}$ as functions of  time. }
\label{timeplots}
\end{figure}

By convention, as in \cite{ISI:000291018000004}, the reconnection rate coefficient is reported as $E_z / B v_A$ where $v_A$ is the Alfv\`en velocity and typical rates associated with fast reconnection are $E_z / B v_A = 0.1 \sim 0.2$. In our simulation, the ions are fixed, and consequently, Alfv\`enic flows are nonexistent.  The characteristic flow velocity for the flux is clearly $\vb{v}_T$. There is a marked increase in the strength of the magnetic field near the reconnection region. Fox et. al. \cite{ISI:000291018000004} account for this effect in the calculation of the local magnetic field, and we perform the same correction. We find that in our simulation, $E_z/B v_T \approx 0.1$, as shown in \cref{timeplots}(a). 

\Cref{timeplots}(b) illustrates the evolution of $\beta$, the ratio of thermal pressure to magnetic pressure.  The sharp peak arises due to rapid heating of the plasma, and then the subsequent decrease comes from the compression of the magnetic field flux before the reconnection process can begin. Once the anisotropic pressure-like term, \cref{recline}(e) supports the out-of-plane electric field, $E_z$, across the reconnection layer, the field compression is maintained and eventually reduced, while the plasma is heated due to the decrease in transport inhibition because of the reconnection process.  This corresponds to the steady increase in $\beta$ as observed after $t=13000\tau_{n}$. \Cref{timeplots}(c) illustrates the magnetization of the plasma over time.  The initial rise in $\omega_c \tau_{ei}$ is due to the compression phase of the magnetic field. After this period, a plateau arises because while the magnetic field decompresses due to reconnection, the plasma heats in the reconnection region, effectively increasing $\tau_{ei}$.  The steady decrease in the late-time behavior is attributed to the magnetic field decompression as the reconnected field lines relax from the Nernst outflows.

When simulations are performed with different values of $\omega_p/\nu_{ei}$ and $v_{\text{th}}/c$, i.e. different plasma density and temperature, they evolve similarly for fixed $\tilde{\delta}_c$, with a reconnection rate stabilizing close to $E_z/B v_T \approx 0.1$, as expected from $H_N$, which only depends on $\tilde{\delta}_c$ for fixed $\omega_c\tau_{ei}$. Hence, the ratio of the skin depth to the collision mean-free-path is  the important consideration for this mechanism. One significant difference with Hall reconnection that we wish to highlight is that quasi-neutrality can be maintained throughout the system and therefore there is no necessity for ion motion outside of the reconnection region to maintain dynamic equilibrium. Redirected heat flows by magnetic reconnection can result in a redistribution of thermal energy and reconnection of field lines can remove thermal transport barriers.  
Since strong heat flows and magnetic fields are expected in the interior of hohlraums, understanding this mechanism can be expected to be important for inertial fusion energy, in particular because reconnection may mitigate the thermal transport inhibition by magnetic fields that could affect the uniformity of the drive. However, the magnetic reconnection could also lead to the production of energetic electrons.

This research was supported by the DOE through Grants No. DE SC0010621 and X and in part through computational resources and services provided by Advanced Research Computing at the University of Michigan, Ann Arbor.

\end{document}